\documentstyle[11pt]{article}
\input{epsf}

\begin{document}
\begin{picture}(150,35)(0,-5)
\put(65,35){\makebox(0,0){\Large \bf
{The seeds of rich galaxy clusters in the Universe} 
}}

\put(65,25){\makebox(0,0){\large F. Governato$^{1}$, C.M. Baugh$^{1}$, 
 C.S. Frenk$^{1}$, S. Cole$^{1}$, C.G. Lacey$^{2}$,  T. Quinn$^{3}$ \& 
J. Stadel$^{3}$}}
\put(65,15){\makebox(0,0){\large 1. Physics Department, Durham University, 
Durham DH1 3LE, UK}}
\put(65,10){\makebox(0,0){\large 2. Theoretical Astrophysics Centre, 
Copenhagen, Denmark.}}
\put(65,5){\makebox(0,0){\large 3. Astronomy Department, 
University of Washington, Seattle, WA 98195}}
\end{picture}

\def\mpc {h^{-1} {\rm{Mpc}}}
\def\and  {\it {et al.} \rm}
\def\rmd {\rm d}

{\bf The discovery\cite{s96} of a population of young galaxies at an
epoch when the universe was about one tenth of its current age has
shed new light on the question of when and how galaxies formed.
Within the context of popular models\cite{bcfl97} this is the
population of primeval galaxies that built themselves up to the size
of present--day galaxies through the process of repeated mergers.  But
the recent detection \cite{s97} of a large concentration of these
primeval galaxies appears to be incompatible with hierarchical
clustering models, which generally predict that clusters of this size
are fully formed later in time.  Here we use a combination of two
powerful theoretical techniques --semi-analytic modelling and N-body
simulations-- to show that such large concentrations should be quite common 
in a universe  dominated by cold dark
matter, and that they  are the progenitors of the rich galaxy 
clusters seen today. We predict the  clustering properties 
 of primeval galaxies which should, when compared with data that will be 
collected in the near future,  test our understanding of galaxy formation
within the framework of a universe dominated by   cold dark
matter.  }


The longstanding search for primeval galaxies was recently brought to
fruition by the combination of deep imaging and Keck Telescope
spectroscopy \cite{sph95}. Because they were originally identified by
a sharp break in their spectrum, at the wavelength corresponding to
the limit of the Lyman series of hydrogen, these galaxies are called
``Lyman break" galaxies. The most distant examples have redshift
$z\simeq 3.5$, corresponding to the time when the universe was only
about 10\% of its current age. The structure discovered by Steidel and
collaborators\cite{s97} appears as a spike in the redshift
distribution at $z=3.09$. It contains 15 Lyman-break galaxies in a
$9\times18$ arcminute field \cite{s97}. The comoving spatial
dimensions of this structure are huge, approximately
$8\times10\times13h^{-1}{\rm Mpc}$ in a Universe with critical
density. (Throughout, we denote Hubble's constant as $H_{0}=100h{\rm
kms}^{-1}{\rm Mpc}^{-1}$.)
\noindent
The current paradigm for the formation of cosmic structure is the cold dark
matter theory \cite{DEFW}. For suitable values of the parameters, this
theory provides a good description of large-scale structure, from scales
probed by microwave background anisotropy measurements to those mapped by
galaxy surveys \cite{sws95}. In this theory, galaxies form by gas cooling
and condensing into dark matter halos which grow by accretion and mergers
in a hierarchical fashion. There is growing observational evidence that
young galaxies are indeed built up from several small pieces
\cite{pasca96}, \cite{low97}, as predicted by the theory.
\noindent 
The technique of semi-analytic modelling provides an effective means
to describe and quantify the complex physics involved in galaxy formation
and evolution. The processes that need to be considered are the
shock heating  and radiative cooling of gas within collapsing dark matter
halos, the subsequent transformation of cold gas into stars, the regulation
of star formation by feedback from stellar winds and supernovae, and the
merging of galaxy fragments. All these processes can be encoded in a set of
simple, physically motivated rules. A surprisingly small number of free
parameters are involved, and these are set by reference to a small subset
of local galaxy data. The result is a fully specified model of galaxy
formation that can be used to make predictions for the properties of the
high redshift universe. Semi-analytic models have been remarkably
successful in reproducing or predicting a range of galaxy properties, such
as the luminosity function, number counts and colours \cite{kwg93}
\cite{cafnz}, and the evolution of the Hubble sequence of morphological
types \cite{bcf96a} \cite{k96}. In particular, predictions for the cosmic
history of star formation \cite{wf} \cite{cafnz}, were subsequently found
to be in good agreement with observations \cite{madau96}.

\begin{figure*}
\begin{picture}(300,600)
\put(20,450)
{\epsfxsize=10.5truecm \epsfysize=12.5truecm 
\epsfbox[0 0 570 700]{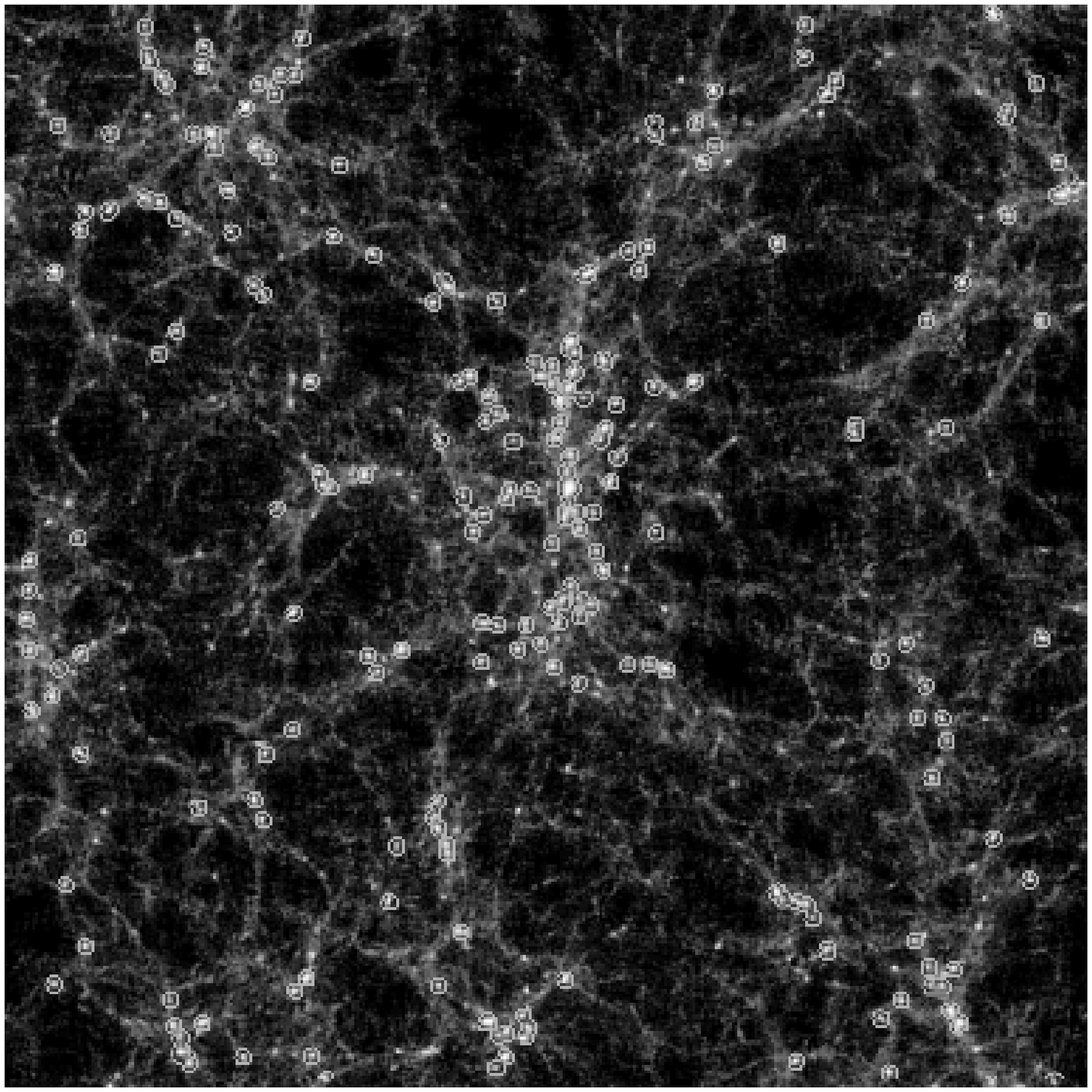}}
\put(20,335)
{\epsfxsize=10.5truecm \epsfysize=12.5truecm 
\epsfbox[0 0 570 700]{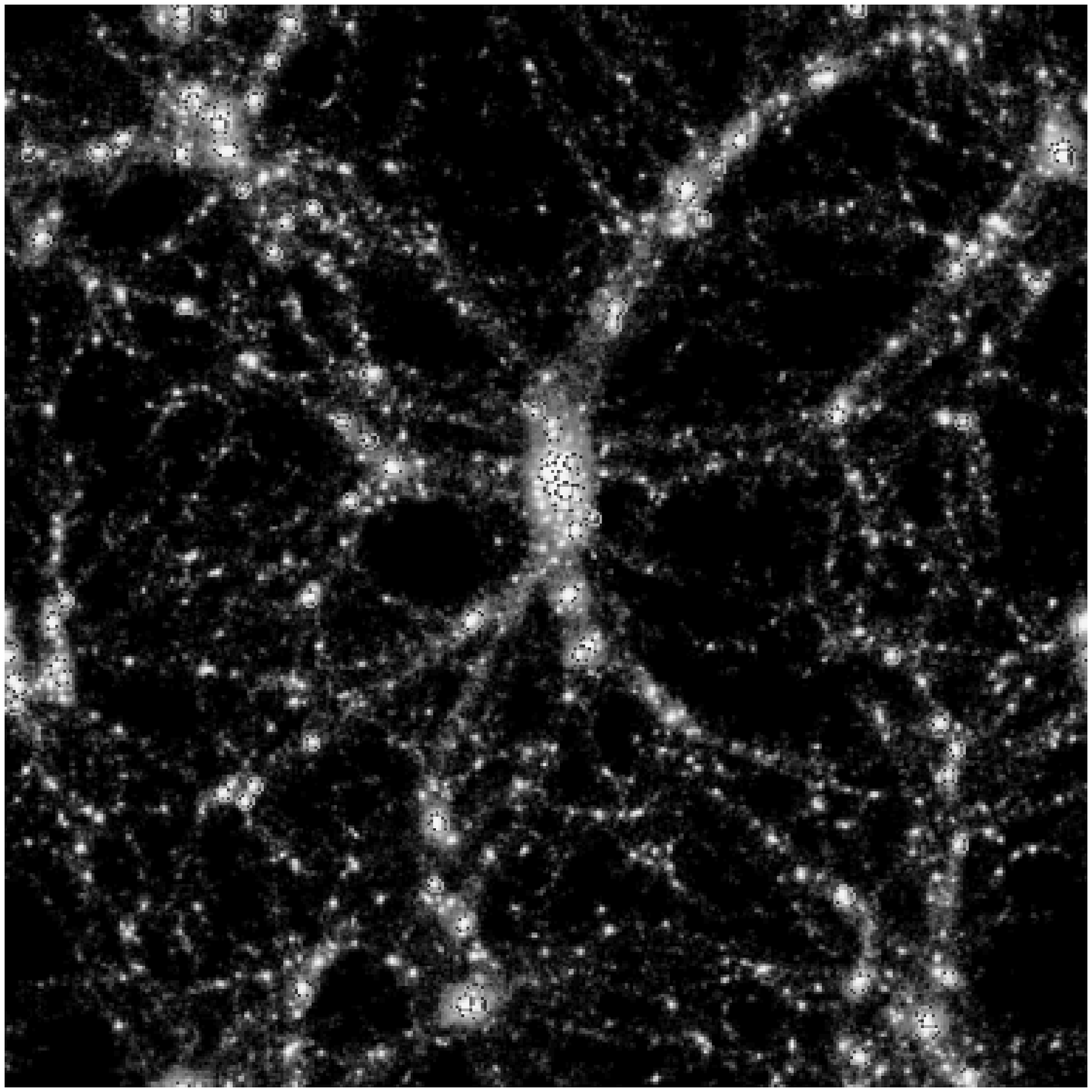}}
\end{picture}
\caption[junk]{}
\label{slice}
\end{figure*}

\begin{figure*}
{\epsfxsize=17.5truecm \epsfysize=24.5truecm 
\epsfbox[100 0 570 700]{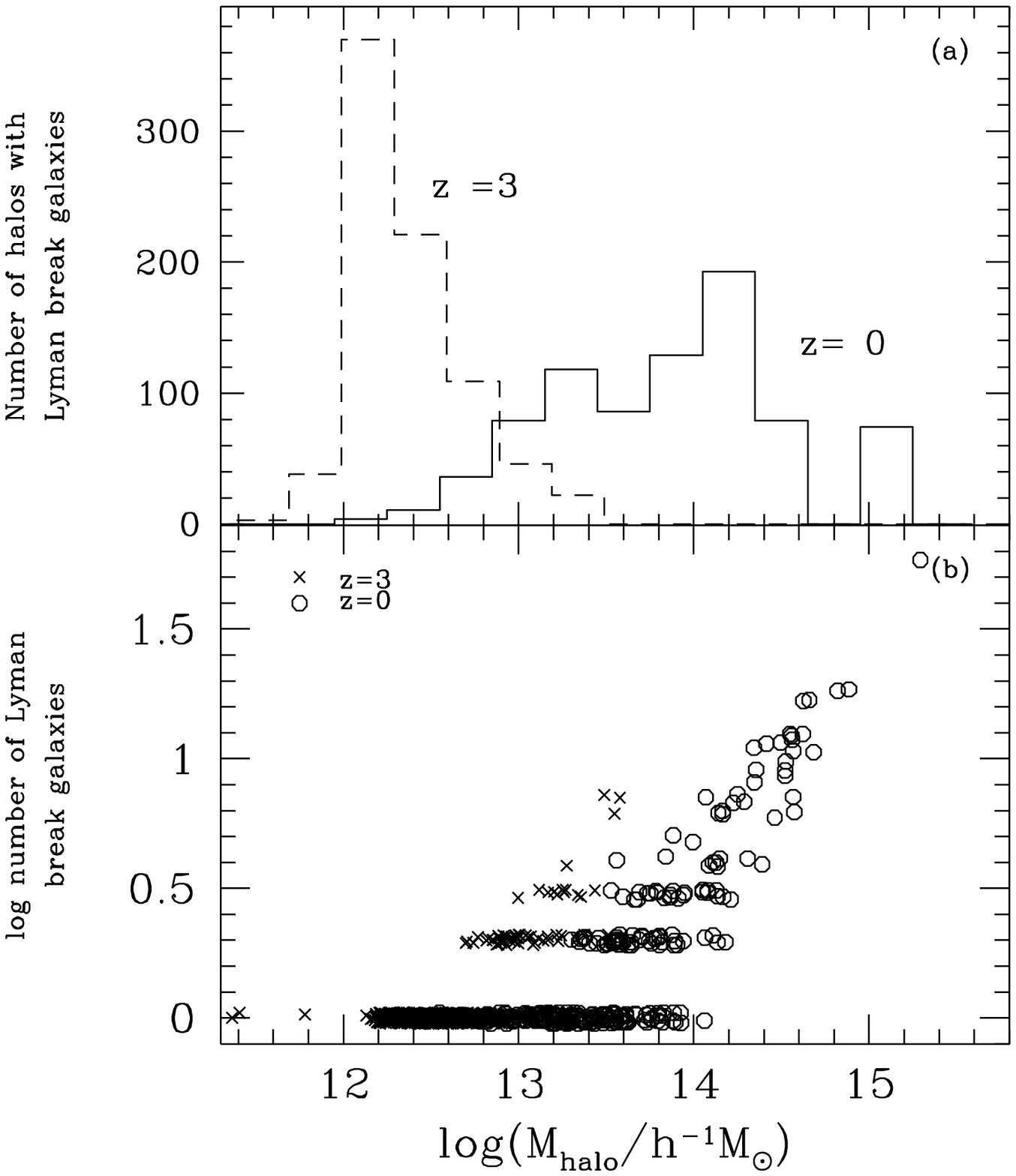}}
\caption[junk]{}
\label{halodist}
\end{figure*}

The semi-analytic approach on its own provides little direct information
about the spatial distribution of galaxies. To overcome this limitation we
applied the semi-analytic galaxy formation rules to dark matter halos grown
in high resolution N-body simulations, an approach previously employed to
study the clustering of the local galaxy population \cite{kns97}. The
supercomputer simulations that we analyzed followed the formation of
structure by gravitational instability in two cold dark matter (CDM)
universes, one with the critical density, $\Omega=1$, and the other with
present day density parameter $\Omega_0=0.3$ (and $h=0.75$). These span the
range of viable models of structure formation of this type.  Dark matter
halos were identified in the simulations at redshift $z=3$ and populated
with visible galaxies according to the predictions of the semi-analytic
model for halos of each mass \cite{bcfl97}. In this way we constructed
catalogues of the positions, velocities and spectrophotometric properties
of all the galaxies present in the simulations, including the effect on the
colours of the galaxies produced by absorption by intervening gas
\cite{madau95}. The model galaxies were then ``observed'' in the same
filters used by Steidel and collaborators, and mock Lyman-break galaxies
were identified in exactly the same way as in the real survey \cite{s97}.
The spatial distribution of the dark matter and Lyman-break galaxies
in our $\Omega=1$ simulation is illustrated in Figure~\ref{slice}. The
top half of the picture shows a slice through the simulation at $z=3$,
$1.6$ Gyr after the Big Bang in this cosmology for
$h=0.5$. Lyman-break galaxies are seen to have formed within the
largest dark matter halos present at that epoch and to trace the
densest ridges of the dark matter distribution.  A redshift survey
covering the central region of this slice would reveal a ``spike" like
that observed by Steidel and collaborators\cite{s97}.  The fate of the
Lyman-break galaxies in this ``spike" is illustrated in the bottom panel 
of Figure~\ref{slice} which shows the same slice, but at the present
day, $11.5$ Gyr later. The material in the region of the spike has
collapsed to form the large cluster seen near the centre of the
image. This cluster has a mass of $9\times10^{14}h^{-1}M_{\odot}$,
comparable to the mass of the Coma cluster \cite{snef}, and contains
the descendants of 60 objects originally identified as Lyman-break
galaxies.
The example in Figure~\ref{slice} illustrates the tendency for
Lyman-break galaxies to form preferentially in large dark matter
halos.  The dashed line in Figure~\ref{halodist}(a) shows the
predicted mass distribution of halos that hosted Lyman-break galaxies
at $z=3$. The solid lines give the corresponding distribution for the
present day halos in which the Lyman-break galaxies ended
up. According to the theory, today's descendants of Lyman-break
galaxies exhibit a marked preference for dense environments, ranging
from poor groups to rich clusters, with only a small fraction
surviving as isolated galaxies. Thus, as conjectured by Steidel and
co-workers\cite{s97}, large concentrations of Lyman-break galaxies at
high redshift pick out regions where proto-groups and proto-clusters
are forming.  Our simulations clearly show that the precursors of
clusters are highly irregular and elongated, reflecting their ongoing
formation by accretion of massive lumps at the intersection of long
filaments. It is rather interesting that the field observed by Steidel
and collaborators\cite{s97} contains a quasar, supporting the view that dense
environments favour the formation of active galaxies \cite{yee}.

With our simulations we are able to calculate the probability of
finding structures as large as the one discovered by Steidel and
coworkers\cite{s97}.  To do this, we first located galaxies in
cylindrical `skewers' of cross-sectional area equal to the area of the
survey, using the simulation volumes and their periodic replications
to sample a unit redshift interval. We then smoothed the galaxy
density field with a top hat filter of width $\Delta z=0.04$, equal to
the binwidth in the actual survey. Finally, we obtained the frequency
of distinct peaks with 3.4 time the mean number of Lyman-break
galaxies, which is the overdensity of the most significant spike in
the real survey.  We find that 45\% of all skewers contain at least
one such spike in a unit redshift interval centred on $z=3$ and a
significant number contain 2 or more.  These numbers are very similar
in the two CDM models we have examined. We conclude therefore that
structures comparable to the spike recently discovered are not at all
surprising in cold dark matter models irrespective of the detailed
values of the cosmological parameters.

\begin{figure*}
{\epsfxsize=16.5truecm \epsfysize=18.5truecm 
\epsfbox[70 160 570 700]{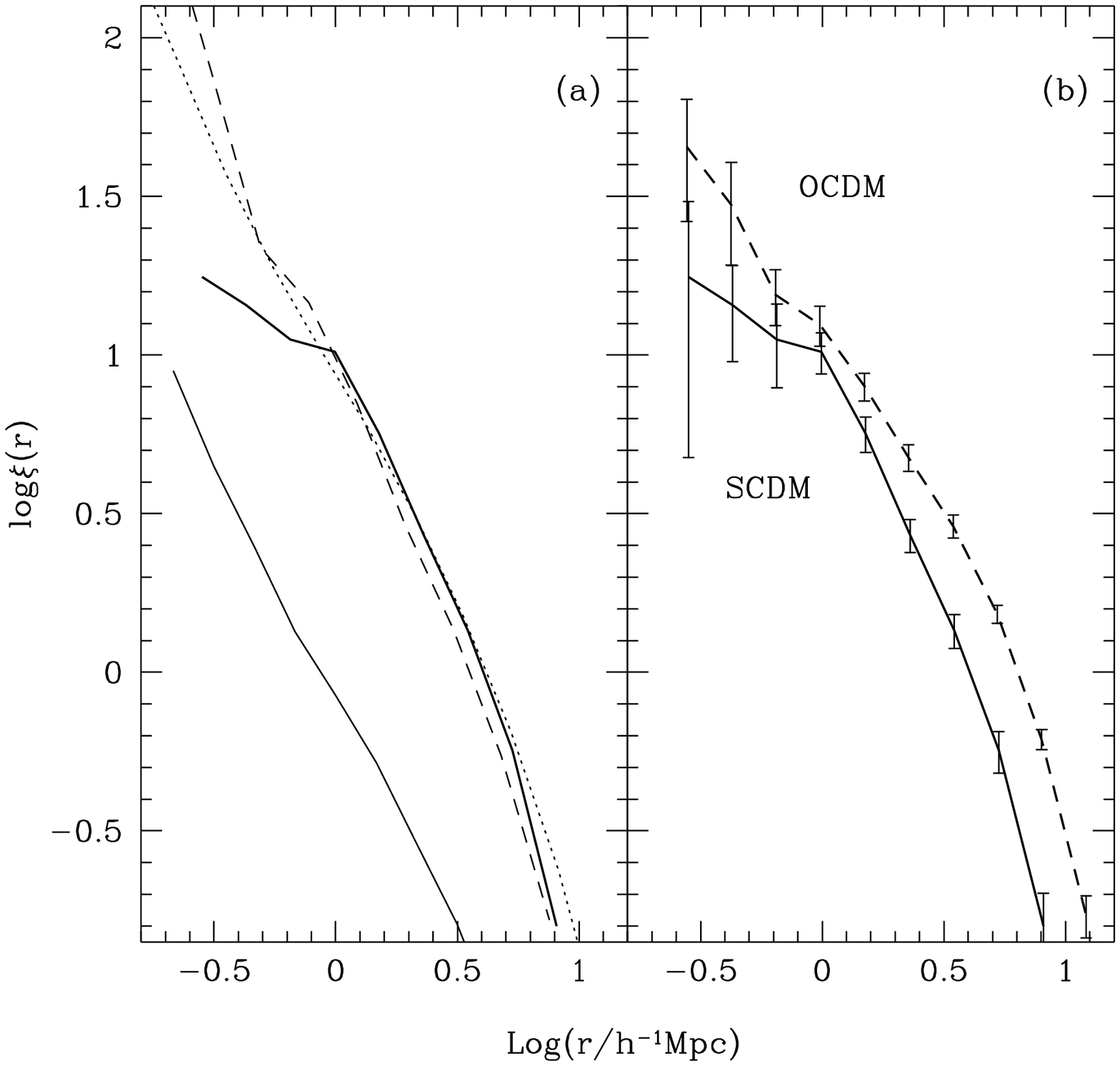}}
\caption[junk]{}
\label{xsi}
\end{figure*}

A more quantitative comparison between theory and observations is provided
by the correlation function of Lyman-break galaxies, a statistic which is,
in principle, measurable from the large surveys currently
underway. Previous theoretical attempts to quantify clustering at high
redshift have been restricted to dark matter halos \cite{mo96},
\cite{bagla97}. With the semi-analytic approach, on the
other hand, we are able to predict directly the clustering properties of
the {\it observable} galaxies themselves. There are substantial differences
between the clustering statistics of dark halos and Lyman-break galaxies
due, for example, to scatter in the relation between halo mass and galaxy
luminosity and to the fact that about 10\% of dark halos contain multiple
Lyman-break galaxies (see Figure~\ref{halodist}(b)). 
Our predicted correlation function for Lyman-break galaxies 
at redshift $z=3$ is plotted in Figure~\ref{xsi}. Its
amplitude is over 10 times larger than that of the underlying mass
distribution, reflecting the preferential formation of Lyman-break galaxies
in dark matter halos of mass $\ge 10^{12}h^{-1}M_{\odot}$ (which are rare
at $z\simeq 3$), and consistent with the analytic predictions of
\cite{bcfl97}. The bias parameter (defined as the ratio of rms fluctuations
in the galaxies and mass respectively in spheres of radius $8h^{-1}{\rm
Mpc}$) is $b=4.2\pm 0.4$ for the $\Omega=1$ model and $b=2.5\pm 0.2$ in the
$\Omega_0=0.3$ model.

Cold dark matter models of galaxy formation predict that only a small
fraction of the stars present today -- perhaps as little as $5\%$ -- have
formed by $z\simeq3$ \cite{cafnz}. In spite of this, the observed abundance
of Lyman-break galaxies can be accounted for in these models \cite{bcfl97},
although the results depend sensitively on the amplitude of density
fluctuations, the choice of stellar initial mass function and the possible
effects of dust on galaxy luminosities. Thus, unless the effects of dust
have been severely underestimated, the discovery of the Lyman-break
galaxies indicates that the formation sites of most of the stars in our
universe have now been identified. According to theory, the Lyman-break
galaxies are the brightest, most massive, examples of the population of
young galaxies present at $z\simeq3$. Because of this, they are expected to
be clustered much more strongly than the underlying dark matter, an
expectation that, as we have shown, is consistent with the detection of a
large spike in their redshift distribution. These data provide
incontrovertible evidence for {\it biased} galaxy formation. At lower
redshifts, other surveys \cite{cfrs96} generally pick out less extreme
representatives of the galaxy population which tend to be less strongly
biased than the Lyman-break galaxies are at $z\simeq3$.  Using the
theoretical approach that we have outlined here it is
possible to relate data on galaxy clustering at various redshifts with
theoretical expectations. Such comparisons, particularly with forthcoming
surveys at high redshift, will test the model  of galaxy formation in a
cold dark matter universe.

\noindent{\bf Acknowledgements.} We acknowledge useful discussions with
Simon White. The simulations were performed at ARSC and NCSA supercomputing
centers. This work was supported by the EU Network for Galaxy Formation and
Evolution, the NASA/ESS programme, and PPARC.  CSF acknowledges a PPARC
Senior Research Fellowship.  CGL was supported by the Danish National
Research Foundation through its establishment of the Theoretical
Astrophysics Center.

\noindent {\bf  Correspondence} and requests for materials should be
 addressed to \\F.G  (Fabio.Governato@durham.ac.uk) and C.S.F 
(c.s.frenk@durham.ac.uk)

\newpage

\section*{Figure Captions}

\setlength{\parindent}{0mm}

{\bf Figure 1.} ({\it a}) 
The distribution of mass and Lyman-break galaxies in a slice of an
$\Omega=1$ CDM simulation at $z=3$.  The slice has side $50h^{-1}{\rm Mpc}$
and thickness $5h^{-1}{\rm Mpc}$. The region plotted contains a large
cluster at $z=0$.  The simulation was performed using a parallel tree-code
with 3 million particles.  The high resolution allows dark matter halos as
small as that of the Large Magellanic Cloud to be resolved.  The initial
amplitude of density fluctuations was chosen so as to approximately
reproduce the observed abundance of rich galaxy clusters today
\cite{vince}.  The logarithm of the dark matter density is colour-coded so
that the highest density regions are white and the lowest density regions
are black.  The blue circles mark the locations where, according to our
semi-analytic galaxy formation model, Lyman-break galaxies have formed.
Whenever an individual dark matter halo contains more than one Lyman-break
galaxy, the first one is placed at the centre of the halo and the rest at
the positions of randomly selected halo particles.  According to the
semi-analytic model, there are $900$ Lyman-break galaxies in the entire
$(50h^{-1}{\rm Mpc})^3$ simulation volume.  As expected from \cite{bcfl97},
both our $\Omega=1$ and $\Omega_0=0.3$ models produce an abundance of
Lyman-break galaxies at high redshift consistent with the observed counts.
The Lyman-break galaxies are preferentially found in the higher density
regions and, as a result, they are strongly {\it biased} relative to the
underlying mass distribution. ({\it b}) The same slice as in ({\it a})
evolved to the present day. The blue circles mark the positions of
particles which were associated with Lyman-break galaxies at $z=3$. The
largest halo in the simulation volume has a mass comparable to that of the
Coma cluster and contains the descendants of 60 Lyman-break galaxies. The
general appearance of the $\Omega_0=0.3$ simulation is qualitatively
similar.

\vskip 1truecm

{\bf Figure 2.} 
(a) The mass distributions of dark matter halos 
that host Lyman-break galaxies at redshift 3 ({\it dashed lines}) 
and of halos that host their descendants at the present time 
({\it solid line}). At high redshift, Lyman-break
galaxies are preferentially located within the most massive halos that have
formed at that time. These halos tend to merge together to form the most
massive structures present at $z=0$. Today, the descendants of Lyman-break
galaxies are to be found in a range of environments, from groups to
clusters, with only a small fraction surviving in smaller, isolated halos.
(b) The number of Lyman break galaxies as a function of 
halo mass at z=3 (crosses). Around $10\%$ of the halos that host Lyman 
break galaxies contain more than one of these objects.
The circles show the number of Lyman break progenitors as a function of 
halo mass at $z=0$. The most massive cluster in the simulation volume 
had $60$ progenitors that were identified as Lyman break galaxies at $z=3$.

{\bf Figure 3.} {\it Left panel:} 
Correlation functions, $\xi(r)$, in the $\Omega=1$ CDM simulation at $z=3$. The
abscissa gives the pair separation in comoving coordinates. The dashed line
shows the true spatial correlation function of the Lyman-break
galaxies. The thick solid line shows the apparent correlation function in
redshift space, -- the correlations are reduced on small scales because of
smearing by random motions within virialized halos and are slightly
enhanced on large scales due to coherent inflow onto collapsing
structures. The dotted line shows the analytic prediction of \cite{bcfl97}
for the real-space correlation function.  Overall, the agreement with the
simulation result is very good. On small scales the simulation results are
slightly higher, reflecting small inaccuracies in the analytic model. On
large scales the simulation results are slightly suppressed by finite
box effects in our relatively small simulation volume. On these large
scales the analytic results are more reliable.  The thin solid line shows
the cold dark matter correlation function at $z=3$. {\it Right panel:} The
redshift space correlation functions in our $\Omega=1$ (SCDM, solid line) and
$\Omega_0=0.3$ (OCDM dashed line) simulations. The error bars were obtained by
bootstrap resampling. The comoving clustering length of Lyman-break
galaxies (defined as the pair separation at which $\xi(r)=1$) at $z=3$ is
comparable to the clustering length of bright galaxies today. Perhaps
counterintuitively, the Lyman-break galaxy correlation functions in the two
models are quite similar, with characteristic clustering lengths which
differ by only $\sim 50\%$. (The inclusion of a non-zero cosmological
constant makes very little difference to the results of an $\Omega_0\simeq
0.3$ model; cf. Fig~8 of \cite{bcfl97}). This is due in part to the
normalization of the simulations which requires that they should
approximately reproduce the observed abundance of clusters at the present
day.

\newpage

\begin{thebibliography}{99}{
\newcommand{\mn}{{\em Mon. Not. R. Astr. Soc., }}
\newcommand{\apj}{{\em Astrophys. J., }}
\newcommand{\aj}{{\em Astron. J., }}
\renewcommand{\aa}{{\em Astr. Astrophys., }}
\newcommand{\ass}{{\em Astrophys. Space Sci., }}
\newcommand{\nat}{{\em Nature, }}

\bibitem{s96} Steidel, C.C., Giavalisco, M., Pettini, M., Dickinson, M., 
                 Adelberger, K. Spectroscopic Confirmation of a Population of
Normal Galaxies at Redshifts $z > 3$.  \apj\, {\bf 462} L22 (1996).
\bibitem{bcfl97} Baugh, C.M., Cole, S., Frenk, C.S., Lacey, C.G.,
 The Epoch of Galaxy Formation.   \apj (in the press)
\bibitem{s97} Steidel, C.C., Adelberger, K.L., Dickinson, M., Giavalisco, M., 
              Pettini, M., Kellogg, M. A large structure of galaxies at z
              $\sim 3$ and its cosmological implications.  \apj {\bf 492}, 428-438
\bibitem{sph95} Steidel, C.C., Pettini, M., Hamilton, D. Lyman imaging of
              high--redshift galaxies. III. New observations of four QSO
              fields. \aj {\bf 110}, 2519-2536 (1995).
\bibitem{DEFW}Davis, M., Efstathiou, G., Frenk, C.S. \& White, S.D.M. The evolution of large--scale structure in a universe dominated by cold dark matter.  \apj 292, 371-394 (1985).
\bibitem{sws95} Scott, D, Silk, J. \& White, M.  
From microwave anisotropies to cosmology {\em Science}, 
{\bf 268}, 829-835 (1995).
\bibitem{pasca96}Pascarelle, S.M., Windhorst, R.A., Keel, W.C., Odewahn, S.C.
Sub-galactic clumps at a redshift of 2.39 and implications for galaxy
 formation.  \nat 383, 45-50 (1996).
\bibitem{low97}Lowenthal, J.D {\it et al.} Keck Spectroscopy for redshift z
 $\sim 3$ galaxies in the Hubble Deep Field. \apj 481, 673-688 (1997).
\bibitem{kwg93} Kauffmann, G., White, S.D.M., Guiderdoni, B. The formation
and evolution of galaxies within merging dark matter haloes.  \mn {\bf
264}, 201-218 (1993).
\bibitem{wf} White, S.D.M., Frenk, C.S. Galaxy formation through hierarchical
clustering.   \apj 379, 52-79 (1991).
\bibitem{cafnz} Cole, S., Aragon-Salamanca, A., Frenk, C.S., Navarro, J.F., 
                 Zepf, S.E. A recipe for galaxy formation.  \mn {\bf
                 271}, 781-806 (1994).
\bibitem{bcf96a} Baugh, C.M., Cole, S., Frenk, C.S. Evolution of the Hubble
                 sequence in hierarchical models for galaxy formation.  \mn
                 {\bf 283}, 1361-1378 (1996).
\bibitem{k96} Kauffmann, G. The age of elliptical galaxies and bulges in a 
merger model. \mn {\bf 281}, 487-492 (1996).
\bibitem{madau96} Madau, P., Ferguson, H.C., Dickinson, M.E., Giavalisco, M.,
                  Steidel, C.C., Fruchter, A. High--redshift galaxies in
                  the Hubble Deep Field: colour selection and star
                  formation history to z $\sim 4$. \mn {\bf 283},
                  1388-1404 (1996).
\bibitem{kns97} Kauffmann, G., Nusser, A., Steinmetz, M. Galaxy formation
                  and large--scale bias  \mn  {\bf 286}, 795-811 (1997).
\bibitem{madau95} Madau, P., Radiative transfer in a clumpy universe: The
colors of high--redshift galaxies.  \apj {\bf 441}, 18-27 (1995).
\bibitem{snef}  White, S.D.M., Navarro J.,F.,Evrard A.E., Frenk C.S. The Baryon content of galaxy clusters -- a challenge to cosmological orthodoxy. 
 \nat {\bf 366}, 429-433 (1993).
\bibitem{mo96}Mo, H.J. \& Fukugita, M. Constraints on the cosmic structure
formation models from early formation of giant galaxies.  \apj {\bf 467} L9-L13. (1996).
\bibitem{bagla97} Bagla, J.S. Evolution of galaxy clustering.  \mn (in press). 
\bibitem{vince} Eke, V.R., Cole, S., Frenk, C.S. Cluster evolution as a
diagnostic for Omega.  \mn {\bf 282}, 263-280 (1996).
\bibitem{cfrs96} Lefevre, O., Hudon, D., Lilly, S. J., Crampton, D., 
Hammer, F., Tresse, L.The Canada-France Redshift Survey: 
VIII. Evolution of the Clustering of Galaxies from $z \approx 1$
\apj 461, 534-545 (1996).
\bibitem{yee}Ellingson, E., Yee, H. K. C., Green, R. F., Kinman, T. D.
Clusters of galaxies associated with quasars. I - 3C 206 \aj {\bf 97},
1539-1549 (1989). 
} 
\end{thebibliography}
\end{document}